\documentstyle[aps,preprint,prc]{revtex}
\def\beq{\begin{equation}}  
\def\eeq{\end{equation}}  
\begin{document}
\draft 

\title{An extension of $RPA$ preserving energy weighted sum rules. \\ 
An application to a 3-level Lipkin model.}

\author{M. Grasso and F. Catara}
\address{Dipartimento di Fisica, Universit\`a di Catania}
\address{and}
\address{Istituto Nazionale di Fisica Nucleare, Sez. di Catania}
\address{Corso Italia 57, I-95129 Catania, Italy}
\maketitle

\begin{abstract}
A limitation common to all 
extensions of $RPA$ including only particle-hole configurations is that they
violate to some extent the Energy Weighted Sum Rules. Considering one such
extension, the improved $RPA$ ($IRPA$), already used to study the electronic
properties of metallic clusters, we show how it can be generalized in order
to eliminate this drawback. This is achieved by enlarging the configuration
space, including also elementary excitations corresponding to the annihilation
of a particle (hole) and the creation of another particle (hole) on the
correlated ground state. The approach is tested within a solvable 3-level
model. \end{abstract}

\section{Introduction}
Collective excitations are a common feature of a large variety of many 
body systems. Their properties are intimately related to the stucture of the 
ground state upon which they are built. The simplest theory of excited states
of a quantum system where correlations are taken into account to some extent
is the Random Phase Approximation ($RPA$). In this theory one
introduces a set of operators $Q^\dag _\nu$, whose action on the ground
state $|\Psi_0 \rangle$ creates the collective excitations, while the ground
state itself is the vacuum for the $Q_\nu$ operators. The latters are defined
as linear superpositions of particle-hole ($ph$) creation and annihilation
operators,  occupied ($h$) and unoccupied ($p$) single particle states being
defined with  respect to the Hartree-Fock ground state $| HF \rangle$. The $X$
and $Y$  coefficients of these linear forms are solutions of equations which
can be  derived by using the equations of motion method \cite{Ro1,RS}. If the
hamiltonian contains one- and two-body terms, the solution of these equations
would imply the evaluation of one- and two-body density matrices. Standard
$RPA$ is obtained by replacing them by those calculated in the uncorrelated
ground state $| HF \rangle$. This approximation introduces a visible
inconsistency since, on one hand, the definition of $|\Psi_0 \rangle$
as the vacuum of the $Q_\nu$ operators is used to
derive the formal equations determining the $X$ and $Y$ amplitudes; while, 
on the other hand, $|HF \rangle$ is used instead of $| \Psi_0 \rangle$ 
in calculating the expectation values appearing in those equations. 

Various attempts have been made to eliminate this 
inconsistency. We quote the pioneering works \cite{Hara,Ro2} where the 
renormalized $RPA$ ($RRPA$) was introduced. The $RRPA$ was applied to study 
the low-lying spectrum and the transition densities of vibrational nuclei
\cite{KAVOCA} and the double beta decay \cite{tosu,sual,raal} more recently. 
A very important contribution to the solution of this problem has been given
in \cite{schuet}, where a general scheme, the self consistent $RPA$ ($SCRPA$), 
was developed (see also \cite{oromana} and references therein). In reference 
\cite{raal} a fully renormalized 
$RPA$ (fully $RRPA$) has been proposed, which shares some 
similarities with the approach we are going to present in this paper.
In \cite{CAPISAGI,CAGRAPISA} it was shown that by 
using the number operator method \cite{Ro3} it is possible to get for the $X$ 
and $Y$ coefficients a closed set of equations having the same form as in 
$RPA$, 
where the density matrices in the correlated ground state are expressed in 
terms of the $X$ and $Y$ coefficients themselves. Thus the equations to
solve are non  linear and their solution requires a big computational effort. 
In order to appreciate how much a better treatment of correlations modifies
the $RPA$ results, in the same paper a simplified version of the approach was
proposed, the $IRPA$, based on the linearization of the
equations of motion. The simplification consists in contracting the 
two-body terms appearing in the commutator of the
hamiltonian with a one body $ph$  operator with respect to the
correlated ground state. In this way only one body density matrices have
to be evaluated: the so obtained
equations are  still non linear, but they are much easier to be solved, since
only the  one-body density matrix appears. When the latter is calculated in
$|HF \rangle $ rather than in the correlated $| \Psi_0 \rangle $,
 $RPA$ is again obtained \cite{Bro}. This approach was applied in 
\cite{CAPISAGI,CAGRAPISA} to the study of the electronic properties of some 
simple metal clusters, obtaining a better description than $RPA$. However, the
 formulation is quite general and its applicability is by no means limited to 
such systems.

A limitation common to all extensions of $RPA$ 
including only $ph$ configurations
is that they violate to some extent the Energy Weighted Sum Rules 
($EWSR$). In
the present paper we will show that this drawback can be eliminated by
enlarging the configuration space, including also those configurations corresponding to the
annihilation of a particle (hole) and the creation of another particle (hole)
on the ground state. This is in the same spirit of  \cite{dusc,scsc}, where, for the first time, the particle-particle and hole-hole configurations were included within the $SCRPA$ approximation.

Very recently a paper \cite{dan} came to our knowledge, where the same problem is tackled and studied within a solvable 4-level model with a separable residual interaction. As we will show below, there are several differences with the present paper:

1) we explicitly show that the $EWSR$ is exactly satisfied when the configuration space is enlarged;

2) by comparison with the exact solutions of the model we can judge about the quality of the results obtained in $IRPA$ and its enlarged version, with respect to the $RPA$ ones;

3) this comparison allows us to point out that, besides the merit of solving the $EWSR$ problem, the approach has the shortcoming that spurious solutions appear. This problem is not discussed in \cite{dan} where, indeed, probably because a separable residual interaction is used, only one collective state is found despite the fact that 3 elementary excitation modes are present in the model. In this context, it is worth mentioning that spurious solutions are also found in 
\cite{raal}, where they are interpreted as "new excitation modes".

The paper is organized as follows. In Section II we shortly recall the main 
$IRPA$ equations, pointing to the origin of the $EWSR$ violations. Then we show
how this problem is solved when the enlarged space is considered. In Section 
III we illustrate the approach by applying it to a solvable 3-level model
\cite{su3,hoyu,samba} and comparing the different approximations among themselves
and with the exact results.

\section{Formulation of the approach}

In this section we 
recall the main steps leading to the $IRPA$ equations, presented in detail in 
\cite{CAPISAGI,CAGRAPISA}, and illustrate why, being
limited to $ph$ excitations, the $IRPA$ approximation   violates the $EWSR$
\cite{RS}. Then we show that, enlarging the space by including also $pp$ and
$hh$ configurations, this difficulty is overcome. In this respect, our
approach is similar to the fully $RRPA$ \cite{raal}.

\subsection{$IRPA$ and the $EWSR$ problem}

Let $| \Psi_0 \rangle $ be the ground state of the system and $|\Psi_{\nu} \rangle$ its excited states. Assuming 
that the latters are linear combinations 
of $ph$ and $hp$ configurations built upon $| \Psi_0 \rangle $ one writes:
\beq
|\Psi_{\nu} \rangle \equiv Q^{\dag}_{\nu} |\Psi_0 \rangle \equiv 
\sum_{ph} [X^{\nu}_{ph} B^{\dag}_{ph} - Y^{\nu}_{ph} B_{ph} ] | \Psi_0 
\rangle~,
\label{1}
\eeq
where 
$p$ ($h$) denotes the quantum numbers of an unoccupied ($p$) and occupied
($h$) single particle state in the uncorrelated Hartree-Fock reference state
$|HF \rangle $. In eq.(\ref{1}) we have introduced renormalized $ph$ creation
($B^\dag$) and annihilation ($B$) operators. In \cite{CAPISAGI,CAGRAPISA} it
is shown that in the basis diagonalizing the one-body density matrix they can
be written as: \beq B^{\dag}_{ph} = D^{-{1\over2}}_{ph} a^{\dag}_p a_h~,
\label{2} \eeq with: \beq
D_{ph} \equiv n_h - n_p~,
\label{3}
\eeq
where $n_h$ and $n_p$ are respectively
 the hole and particle occupation numbers in the correlated ground state $|
\Psi_0 \rangle$. Assuming that $|\Psi_0 \rangle$ is the vacuum of the $Q_\nu$
operators, \beq Q_\nu | \Psi_0 \rangle = 0~,
\label{4}
\eeq
the ortonormality conditions for the excited states leads to:
\beq
\delta_{\nu \nu'}=\langle \Psi_{\nu} | \Psi_{\nu'} \rangle=
\sum_{ph} [X^{\nu *}_{ph} X^{\nu'}_{ph}-Y^{\nu *}_{ph} Y^{\nu'}_{ph} ]~.
\label{5}
\eeq
The equations determining 
the $X^\nu$ and $Y^\nu$ amplitudes and the excitation energies $E_\nu$ of the
states $|\Psi_{\nu} \rangle$ are obtained by using the equations of motion
method \cite{Ro1,RS}. They read: \beq \left(\begin{array}{cc}
A&B\\
B^*&A^* \end{array}
\right )\left(\begin{array}{c}X^{\nu}\\Y^{\nu}\end{array}\right )=
E_{\nu}\left (\begin{array}{c}X^{\nu}\\ -Y^{\nu}\end{array}\right )~,
\label{6}
\eeq
with the $A$ and $B$ matrices given by:
\beq
A_{ph,p'h'} = \langle \Psi_0 |[B_{ph},H,B^{\dag}_{p'h'}|\Psi_0\rangle
\label{7}
\eeq
and
\beq 
B_{ph,p'h'} = -\langle \Psi_0 |[B^{\dag}_{ph},H,B^{\dag}_{p'h'}]|\Psi_0
\rangle~. \label{8}
\eeq
In (\ref{7}) and (\ref{8}) $H$ is the hamiltonian of the system and:
\beq
[A,B,C] \equiv {1\over2} ([A,[B,C]]+[[A,B],C])
\label{9}
\eeq
The standard $RPA$ equations can be obtained 
by putting $n_h$=1, $n_p$=0 in the expressions for the operators $B$ and
$B^{\dag}$ (\ref{2}) and by replacing the correlated ground state $| \Psi_0
\rangle$ appearing in  (\ref{7}) and (\ref{8}) with the Hartree-Fock one $|HF
\rangle$. In \cite{Bro} it is shown that the 
$RPA$ equations can equivalently be obtained by
linearizing the commutator $[H,B^{\dag}_{p'h'}]$ in (\ref{7}) and (\ref{8}),
i.e. by contracting it with respect to $|HF \rangle$. A better approximation
is done in $IRPA$, where the linearization is made by contraction in $|\Psi_0
\rangle$. In a loose notation, this means: \begin{eqnarray}
\nonumber
[H,a^{\dag} _p a_h ] \rightarrow a^{\dag} a + a^{\dag} a^{\dag} a a  \\
\sim a^{\dag} a + \langle \Psi_0 | a^{\dag} a | \Psi_0 \rangle a^{\dag} a ~.
\label{10}
\end{eqnarray}
Therefore, the occupation numbers in the correlated ground state appear in the
$IRPA$ expressions, while those in $|HF \rangle$ (i.e. 0 or 1) appear in
standard $RPA$.  
This procedure leads to:  
\begin{eqnarray} 
\nonumber
&&A_{ph,p'h'}={1\over2}(D^{1/2}_{ph}D^{-1/2}_{p'h'}+ D^{1/2}_{p'h'}
D^{-1/2}_{ph})   (\epsilon _{p'p}
\delta_{hh'}- \epsilon _{hh'} \delta_{pp'} ) \\ 
&&+D^{1/2}_{ph}D^{1/2}_{p'h'} (hp'|H_2|ph') ~,
\label{11}
\end{eqnarray}
where:
\beq
\epsilon_{p'p} \equiv (p'|H_1|p)+
\sum_{\alpha}n_\alpha (p' \alpha|H_2|p \alpha)
\label{12}
\eeq
and
\beq
\epsilon_{hh'} \equiv (h|H_1|h')  
+\sum_{\alpha}n_{\alpha} ( \alpha h |H_2| \alpha h') ~.
\label{13}
\eeq
For the matrix $B$ one gets:
\begin{equation}
B_{ph,p'h'}=D^{1/2}_{ph}D^{1/2}_{p'h'} (hh'|H_2|pp')~.
\label{14}
\end{equation}
In the above equations $H_1$ is the one-body term of the hamiltonian and $H_2$
its two-body part. We denote by $\alpha$ a generic single particle state
(occupied or unoccupied in $|HF \rangle$). As shown in Appendix A of
\cite{CAPISAGI}, using the number operator method \cite{Ro3} the occupation
numbers appearing in the $A$ and $B$ matrices can be expressed in terms of the
$X$ and $Y$ amplitudes as: 
\beq
n_p=\sum_{h\nu\nu'}(\delta_{\nu\nu'}-{1\over2}\sum_{p_1h_1}
D_{p_1h_1}X^{\nu'}_{p_1h_1}X^{\nu*}_{p_1h_1})D_{ph}Y^{\nu}_{ph}Y^{\nu'*}_{ph}
~,  
\label{15} 
\eeq 
\beq
n_h=1-\sum_{p\nu\nu'}(\delta_{\nu\nu'}-{1\over2}\sum_{p_1h_1}
D_{p_1h_1}X^{\nu'}_{p_1h_1}X^{\nu*}_{p_1h_1})D_{ph}Y^{\nu}_{ph}Y^{\nu'*}_{ph}
~. 
\label{16}
\eeq
Therefore eq.s (\ref{6}) are non linear. 
They have been solved iteratively in the case of metallic clusters
\cite{CAPISAGI,CAGRAPISA}. It is, however, apparent that the approach is quite
general and can be applied to any many body system. The matrices $A$ and $B$
in $IRPA$, eq.s(\ref{11}) and (\ref{14}), are different from those in
standard $RPA$. On one side the Hartree-Fock single particle energies
appearing in the $A$ matrix of $RPA$ are replaced by the quantities appearing
in the first line of eq.(\ref{11}). On the other side, the residual 
interaction in the expressions for $A$ and $B$ is now renormalized by the
factors $D^{1/2}$'s. In $RRPA$ only the latter modification is present. This
latter modification is present also in $RRPA$.

A serious problem arises with respect to the $EWSR$.
 As it is well known, if 
$| \Psi_0 \rangle$ and $| \Psi_{\nu} \rangle$ are a complete set of exact
eigenstates of the hamiltonian, with eigenvalues $E_0$ and $E_\nu$, the
following identity holds: 
\beq 
\sum_{\nu}(E_\nu - E_0) |\langle \Psi_\nu
|F|\Psi_0 \rangle |^2 = {1\over2} \langle \Psi_0 |[F,[H,F]]| \Psi_0 \rangle ~,
\label{17}
\eeq
where $F$ is 
any hermitian single particle operator. The equality (\ref{17}) is in general
violated to some extent when $|\Psi_0 \rangle$, $|\Psi_{\nu} \rangle$ and
$E_{\nu}$ are calculated within some approximation. To which extent it is
satisfied is a measure of the adequacy of the approximation. 
A very important feature of $RPA$
 is that eq.(\ref{17}) is satisfied for any one-body operator if, in
calculating its two sides, one considers $|HF \rangle$ instead of $|\Psi_0
\rangle$ and the solutions of $RPA$ for $| \Psi_{\nu} \rangle$ and $(E_\nu
-E_0)$ \cite{thou}.
This feature follows from the fact that, when
$|HF \rangle$ is used in (\ref{17}) instead of $|\Psi_0 \rangle$ only
particle-hole matrix elements remain in the r.h.s. It is easy to show
\cite{CAPISAGI} that, if the transition operator $F$ has only $p-h$ matrix
elements, the two sides of eq.(\ref{17}) are equal also within $IRPA$.
However, this is not the case in general.

Let us consider separately the two sides of eq.(\ref{17}) in $IRPA$,
 with a general one-body hermitian operator $F$:
\beq
F=\sum_{\alpha \beta} f_{\alpha \beta} a^{\dag}_{\alpha} a_{\beta} ~.
\label{18}
\eeq
The l.h.s. is easily calculated and gives:
\begin{eqnarray}
\nonumber
&&\sum_{\nu}(E_{\nu}-E_0) |\langle \Psi_{\nu}|F|\Psi_0 \rangle|^2 = 
\sum_{\nu}(E_{\nu}-E_0) |\langle \Psi_0 |Q_\nu F|\Psi_0 \rangle |^2 \\
\nonumber
&&=\sum_{\nu}(E_{\nu}-E_0) |\langle \Psi_0 | 
[Q_{\nu},F]|\Psi_0\rangle|^2  \\
&&=\sum_{\nu}(E_\nu -E_0) |\sum_{ph} f_{ph} D^{1/2}_{ph} (X^{\nu}_{ph}+
Y^{\nu}_{ph})|^2 ~,
\label{19}
\end{eqnarray}
which is formally equal to the $RPA$ 
result, apart from the factor $D^{1/2}_{ph}$. Therefore, only the $ph$
components of $F$ enter.  This is due to the fact that the excited states are
described as superpositions of $ph$ configurations only.  Starting from
eq.s(\ref{6}) and using the properties of the $X$ and $Y$ amplitudes,
eq.(\ref{19}) can be written as: 
\begin{eqnarray} 
\nonumber
&&\sum_{\nu} (E_{\nu}-E_0)|\langle \Psi_{\nu} |F|\Psi_0 \rangle |^2  \\
&&=\sum_{ph}D^{1/2}_{ph} f_{ph} \sum_{p'h'} D^{1/2}_{p'h'}f_{p'h'}
(A_{ph,p'h'} - B_{ph,p'h'})~.
\label{20}
\end{eqnarray}
In order to evaluate the r.h.s. of eq.(\ref{17}) one can use for the
commutator $[H,F]$ the same linearization procedure already used in deriving
eq.s(\ref{11}) and  (\ref{14}). It is easy to realize that the result of such
calculation cannot be equal to (\ref{20}) since not only the
$ph$ matrix elements of the residual interaction will appear in it, but also
other terms if they are present in the one body operator $F$. This happens
because the expectation value of the double commutator is taken in the
correlated ground state $|\Psi_0 \rangle$.  We will
show this in the next subsection, where an enlarged configuration space,
including also $pp$ and $hh$ components, will be used to express the excited
states. Of course, if the correlations present in $|\Psi_0 \rangle$ are small
and the occupation numbers do not differ too much from 0 and 1, the violations
of the $EWSR$ are small. But, in general, this is not the case. For example,
for Na clusters, the discrepancy was found \cite{CAPISAGI,CAGRAPISA} to be
about 25$\%$.

\subsection{The enlarged space}

As shown in the previous subsection, the problem of violations of the $EWSR$
 arises because also in $IRPA$, as in $RPA$,
 the excited states 
are expressed as superpositions of $ph$ configurations. Let us then consider
the more general expansion: 
\beq
|{\bar \Psi}_{\nu} \rangle = {\bar Q^{\dag}}_{\nu} |{\bar \Psi}_0 \rangle = 
\sum_{\alpha > \beta} ({\bar X}^{\nu}_{\alpha \beta} B^{\dag}_{\alpha \beta} -
{\bar Y}^{\nu}_{\alpha \beta} B_{\alpha \beta}) |{\bar \Psi}_0 \rangle~,
\label{21}
\eeq
where $\alpha$ and $\beta$ 
stand for any single particle state and $\alpha > \beta$ means that we order
these states according to decreasing occupation numbers, i.e. $n_\alpha <
n_\beta$. The operators $B^{\dag}_{\alpha \beta}$ and $B_{\alpha \beta}$ are
an obvious generalization of eq.s (\ref{2}) and (\ref{3}). As before we define
$|{\bar \Psi}_0 \rangle$ as the vacuum of the ${\bar Q}_{\nu}$ operators: 
\beq
{\bar Q}_{\nu} |{\bar \Psi}_0 \rangle = 0~. 
\label{22} 
\eeq 
In order to make
simpler the notation, we will omit the bars in the collective operators and in
the states, which, of course, are different from those considered in $IRPA$ 
since now $pp$ and $hh$ configurations are included, in addition to the $ph$
ones. The derivation of the equations of motion can be done by following the
same linearization procedure as before. They have the same form as in
eq.(\ref{6}), the matrices $A$ and $B$ being now: 
\begin{eqnarray} 
\nonumber
A_{\alpha \beta ,\gamma \delta}&&={1\over2}(D^{1/2}_{\alpha \beta} 
D^{-1/2}_{\gamma \delta} + D^{1/2}_{\gamma \delta} D^{-1/2}_{\alpha \beta})
(\epsilon_{\alpha \gamma} \delta_{\beta \delta}-\epsilon_{\beta \delta}
\delta_{\alpha \gamma})  \\ &&+D^{1/2}_{\alpha \beta} D^{1/2}_{\gamma \delta} 
(\beta \gamma | H_2| \alpha \delta) 
\label{23} 
\end{eqnarray} 
and
\begin{eqnarray}
\nonumber
B_{\alpha \beta, \gamma \delta}=&&{1\over2}(D^{1/2}_{\alpha \beta} 
D^{-1/2}_{\gamma \delta}-D^{1/2}_{\gamma \delta} D^{-1/2}_{\alpha \beta})
(\epsilon_{\alpha \delta} \delta_{\beta \gamma}-\epsilon_{\beta \gamma}
\delta_{\alpha \delta}) \\
&&+D^{1/2}_{\alpha \beta} D^{1/2}_{\gamma \delta} (\delta \beta|H_2|\gamma \alpha)~,
\label{24}
\end{eqnarray}
where:
\beq
\epsilon_{\alpha \beta}= (\alpha|H_1|\beta)+\sum_{\gamma}n_{\gamma} (\alpha \gamma |H_2|\beta \gamma)~.
\label{25}
\eeq
Apart from the fact that in eq.s(\ref{23}) and (\ref{24}) the indices run
over all single particle states, the main difference with eq.s(\ref{11}) and
(\ref{14}) is the presence of the $\epsilon$ terms also in the $B$ matrix.

Coming back to the $EWSR$ problem, eq.(\ref{20}) is easily generalized to:
\begin{eqnarray}
\nonumber
&&\sum_{\nu}(E_\nu - E_0)|\langle  \Psi_{\nu}|F|\Psi_0 \rangle|^2  \\
&&=\sum_{\alpha > \beta} f_{\alpha \beta} D^{1/2}_{\alpha \beta}
\sum_{\gamma > \delta} f_{\gamma \delta} D^{1/2}_{\gamma \delta} (A_{\alpha
\beta,\gamma \delta} - B_{\alpha \beta, \gamma \delta})~, 
\label{26}
\end{eqnarray}
which, after some tedious manipulations, can be written as:
\begin{eqnarray}
\nonumber
&&\sum_{\nu}(E_\nu -E_0)|\langle \Psi_{\nu}|F|\Psi_0 \rangle|^2  \\
&&={1\over 2} \sum_{\alpha \beta} f_{\alpha \beta}D_{\alpha \beta} \sum_{\gamma \delta} f_{\gamma \delta} [\epsilon_{\alpha \gamma} \delta_{\beta \delta} - \epsilon_{\beta \delta} \delta_{\alpha \gamma} + (\beta \gamma|H_2|\alpha \delta) D_{\gamma \delta}]~.
\label{27}
\end{eqnarray}
The double commutator 
is easily calculated by using the same linearization procedure adopted to
derive the equations of motion. Doing that one realizes that eq.(\ref{17}) is
indeed satisfied. Thus one obtains a kind of generalization of the Thouless
theorem. Namely, eq.(\ref{17}) is satisfied if one calculates its two sides by
using the solutions of the equations of motion and by making the same
approximations introduced in the derivation of the latters.

In principle the new approach 
does not appear to be more difficult than $IRPA$ and its equations can be
solved in realistic cases by the same iterative procedure used there. In
practice, however, the computational effort is much heavier since the
configuration space is much larger. For this reason we have decided to apply
it to a solvable 3-level model \cite{su3,hoyu,samba}.

We show this application in the next section, where we compare the results of 
$IRPA$
 and of its enlarged version with the exact solutions of the model.

\section{The model and the results}

Let us first of all illustrate the solvable model to which we applied the enlarged version of $IRPA$.

It consists 
of three levels, 0, 1 and 2, with energies $\epsilon_0$, $\epsilon_1$ and
$\epsilon_2$, respectively. Let $2\Omega$ be the  degeneracy 
of each level and $N = 2\Omega$ the total number of fermions in the system. 

We define the operators:
\beq
K_{ij} \equiv\sum_m a^{\dag}_{im} a_{jm}~,
\label{28}
\eeq
where the indices $i$ and $j$ 
denote one of the three levels and the index $m$ runs over the $2\Omega$
substates of each of them.  The operators $K$ satisfy the following
commutation relations: 
\beq
[K_{ij},K_{kl}]=\delta_{jk} K_{il} - \delta_{il} K_{kj}~.
\label{29}
\eeq
They are therefore the generators of the $U(3)$ algebra. 
The algebra becomes $SU(3)$ if we consider the additional relation,
\beq
N=\sum_i K_{ii}~,
\label{30}
\eeq
that fixes the total number of particles.

We introduce the hamiltonian for our system as follows:
\begin{eqnarray}
\nonumber
H&&=\sum_{i \neq 0} \epsilon_i K_{ii} + V_0 \sum_{i,j\ne 0} K_{i0} K_{0j} +
V_1 \sum_{i,j \ne 0} (K_{i0} K_{j0} + K_{0j} K_{0i})  \\
&&+V_2 \sum_{i,j,k \ne 0} (K_{i0} K_{jk} + K_{kj} K_{0i})+
V_3 \sum_{i,j,k,l \ne 0}K_{ij} K_{kl}~.
\label{31}
\end{eqnarray}
The terms with the $V_0$ and $V_1$ strengths describe, respectively,
 the $phph$ and $pphh$ 
parts of the interaction. The term with the $V_2$ strength is related to the
$ppph$ part, while the last term represents the $pppp$ part. In standard $RPA$
the only two-body terms of $H$ that enter in the  expressions for the 
matrices $A$ and $B$ of the equations of motion (\ref{6}) are those with the
strengths $V_0$ and $V_1$, i.e. the $ph$ two-body terms. In the $IRPA$
approach \cite{CAPISAGI,CAGRAPISA} the ground state that is actually used in
the calculations is correlated; so the single particle occupation numbers
aren't strictly 1 for hole states and 0 for particle states, as in
$|HF\rangle$. In this case also the $pppp$ term enters in the expressions for
the matrices (actually only in the matrix $A$). In the $IRPA$ approach with
the enlarged configuration space all the terms contribute.

The exact results for the system can be obtained either 
by using the $SU(3)$ symmetry of 
the model or by diagonalizing the hamiltonian (\ref{31}) in the complete set
of states: 
\beq
|n_1 n_2 \rangle \equiv C (K_{10})^{n_1} (K_{20})^{n_2} |0 \rangle~,
\label{32}
\eeq
where $|0\rangle$ denotes the state in which 
all the particles are in the level $0$, $n_1$ and $n_2$ are the numbers of
particles in the levels $1$ and $2$, respectively, and $C$ represents a
normalization factor.

With the same set of parameters chosen for the 
exact calculation, after having performed 
a standard $RPA$ calculation, we solved the equations of motion 
both in the $IRPA$ approach of \cite{CAPISAGI,CAGRAPISA} and 
in the new approach, with the enlarged configuration space.
In the $IRPA$ case 
the operators $Q^{\dag}_\nu$  are defined as linear combinations of $ph$
($i0$, with: $i \ne 0$) and $hp$ ($0i$, with $i \ne 0$) configurations, as in
eq.(\ref{1}): 
\beq 
Q^{\dag}_\nu \equiv \sum_i (X^{\nu}_i {\tilde
K}_{i0}-Y^{\nu}_i{\tilde K}_{0i})~, 
\label{33}
\eeq
while in the enlarged calculation they are defined as in eq.(\ref{21}):
\beq
Q^{\dag}_{\nu} \equiv \sum_{i>j} 
(X^{\nu}_{ij} {\tilde K}_{ij} - Y^{\nu}_{ij} {\tilde K}_{ji})~,
\label{34}
\eeq
where:
\beq
{\tilde K}_{ij} \equiv {1\over{(2\Omega)^{1/2}}} D^{-1/2}_{ij} K_{ij}~.
\label{35}
\eeq
With the definition (\ref{35}) we get, for the excited states of the system,
 the same orthonormality conditions as given in eq.(\ref{5}).

Note that in (\ref{34}) the indices $i$ and $j$ run over all the 3 single
particle levels of the model. In both cases (\ref{33}) and (\ref{34}) we have
solved the non linear problem of eq.s(\ref{6}) by means of an iterative
procedure. We fixed the number of particles $N$ equal to 10. In this case the
number of exact eigenstates of the hamiltonian is 66. The $RPA$ and $IRPA$
calculations will give two excited states, since their configuration space is
composed  only by the two configurations $(1,0)$ and $(2,0)$. The enlarged
$IRPA$ will give three states, since its configuration space is made by the
three configurations $(1,0)$, $(2,1)$ and $(2,0)$. 

We tested various values for the four parameters $V_0$, $V_1$, $V_2$ 
and $V_3$ and for the energies of the levels, the results being qualitatively
the same. In Fig. 1 we show one case, where: 
\beq
\epsilon_0 =0, \hspace{.2in} \epsilon_1 = \epsilon, 
\hspace{.2in} \epsilon_2 = 2.5\epsilon ~, 
\label{36}
\eeq 
\beq
V_0=-\chi, \hspace{.2in} V_1=\chi, \hspace{.2in} V_2={-\chi\over2}  
\hspace{.2in} V_3={\chi\over{10}} ~.
\label{37}
\eeq
Both the $\epsilon$ and $\chi$ parameters have the units of an energy.
In the figure the excitation energies are represented versus the 
increasing values of the strength $\chi$. Dashed lines refer to exact values
of energies. Among all the 66 exact eigenvalues the two represented ones are
those with energies equal to 1 and 2.5 at $\chi = 0$; i.e. those which
correspond to the two $RPA$ and $IRPA$ excited states. Dotted lines refer to
$RPA$ and dot-dashed lines to $IRPA$ values. The three values corresponding to
the enlarged $IRPA$ approach are represented by full lines.

The collapse point of $RPA$, where 
its first excitation energy becomes imaginary, appears at $\chi = 0.024$.
We observe that both the $IRPA$ and 
the enlarged $IRPA$ calculations push the collapse point towards greater
values of the strength parameter $\chi$; so, in this regard, 
both methods improve the $RPA$ results.
Moreover it can be seen 
that the two exact values are better approximated by the first and the third
states obtained in the enlarged $IRPA$ approach, than by the two $IRPA$
states. This is especially evident for the higher state.

It is interesting to focus the attention 
on the presence of the additional state that the enlarged $IRPA$ gives, with
respect to $RPA$ and $IRPA$. Actually this state doesn't correspond to any of
the found exact states.  This fact seems to indicate that it is a spurious
state.  On the other hand its energy is not zero or very small, as it happens
normally for spurious states. Its energy always starts, when $\chi$ starts
from zero, from the energy difference between the levels 1 and 2, and so it
depends on how we fix the values $\epsilon_1$ and $\epsilon_2$. This would
mean that, if we applied our approach in a realistic calculation,  the
spurious states that would appear wouldn't be easily recognized and
eliminated, not having in principle zero or very small energies. This could
cause problems in the interpretation of the calculated spectrum of
excitations.  A similar situation is encountered in $RPA$ at finite
temperature and was also found in \cite{raal}.  Let us look at the transition
probabilities related to this state. Figure 2 shows the transition
probabilities related to the obtained states,for $RPA$ and  $EIRPA$
calculations,  for different $\chi$ strengths.  Let's observe in the figure
the transition probability related to the spurios state,  
\beq 
P_{sp}=|
\langle \Psi_{\nu_{sp}} | F | \Psi_0 \rangle |^2 ~, 
\label{38} 
\eeq   
where
$|\Psi_0 \rangle$ is the ground state, $|\Psi_{\nu_{sp}} \rangle$ is the
spurious state and $F$ the one body operator (\ref{18}) with all the
$f_{\alpha \beta}$'s equal to one. We can see that $P_{sp}$ is very small,
with respect to the other two transition probabilities, only when $\chi$ is
far from the collapse point; when $\chi$ approaches the collapse point the
transition probability (\ref{38}) becomes appreciable (see the case $\chi =
0.04$ in the figure). The same trend is found for other sets of parameters.
   
This means that in the evaluation of any physical quantity, in a realistic
calculation, the existence of spurious states would have some influence and it
would be important to recognize and eliminate them from the calculation. The
problem of how to recognize them is still open, as they don't have in general
small energies and/or small transition probabilities.    

We present now, in
table 1, the results obtained for the $EWSR$, in $IRPA$ and in enlarged $IRPA$
cases. These results refer again to the choice (\ref{36}) and  (\ref{37}) for
the parameters. The violation of the $EWSR$, in the $IRPA$ case, increases
with the increasing values of the $\chi$ parameter and is about of 30$\%$ for
$\chi=0.03$. In the enlarged $IRPA$ case the $EWSR$ is always exactly
satisfied, as expected.

This is an important achievement of the present method, because, as 
stressed in Section 1, all the methods that have been proposed so far in
order to go beyond the $RPA$, by avoiding the inconsistency of the Quasi Boson
Approximation, always violate the $EWSR$ identity.                    

\section{Conclusions} 
In conclusion, we have presented an extension of $RPA$
which avoids the use of the Quasi Boson Approximation and, at variance with
many other attempts made in the same direction, preserves exactly the $EWSR$.
This is obtained as a generalization of a previously studied approach by
enlarging the configuration space with respect to that commonly used, which
contains only particle-hole elementary escitations.  The approach has been
tested on a 3-level solvable model.

The authors gratefully thank P. Schuck and M. Sambataro for helpful discussions.

\newpage

\begin{figure}
\caption{Excitation energies versus ${\chi \over \epsilon}$, with the 
(\ref{36})-(\ref{37})   parameters. The energies in $Y$ axis are expressed in
units of $\epsilon$.} \label{Fig. 1}
\end{figure}

\begin{figure}
\caption{Transition probabilities 
for four values of ${\chi \over \epsilon}$, with the (\ref{36})-(\ref{37})
parameters. The energies in $X$ axis are expressed in units of $\epsilon$.}
\label{Fig. 2} \end{figure}

\newpage
\includegraphics{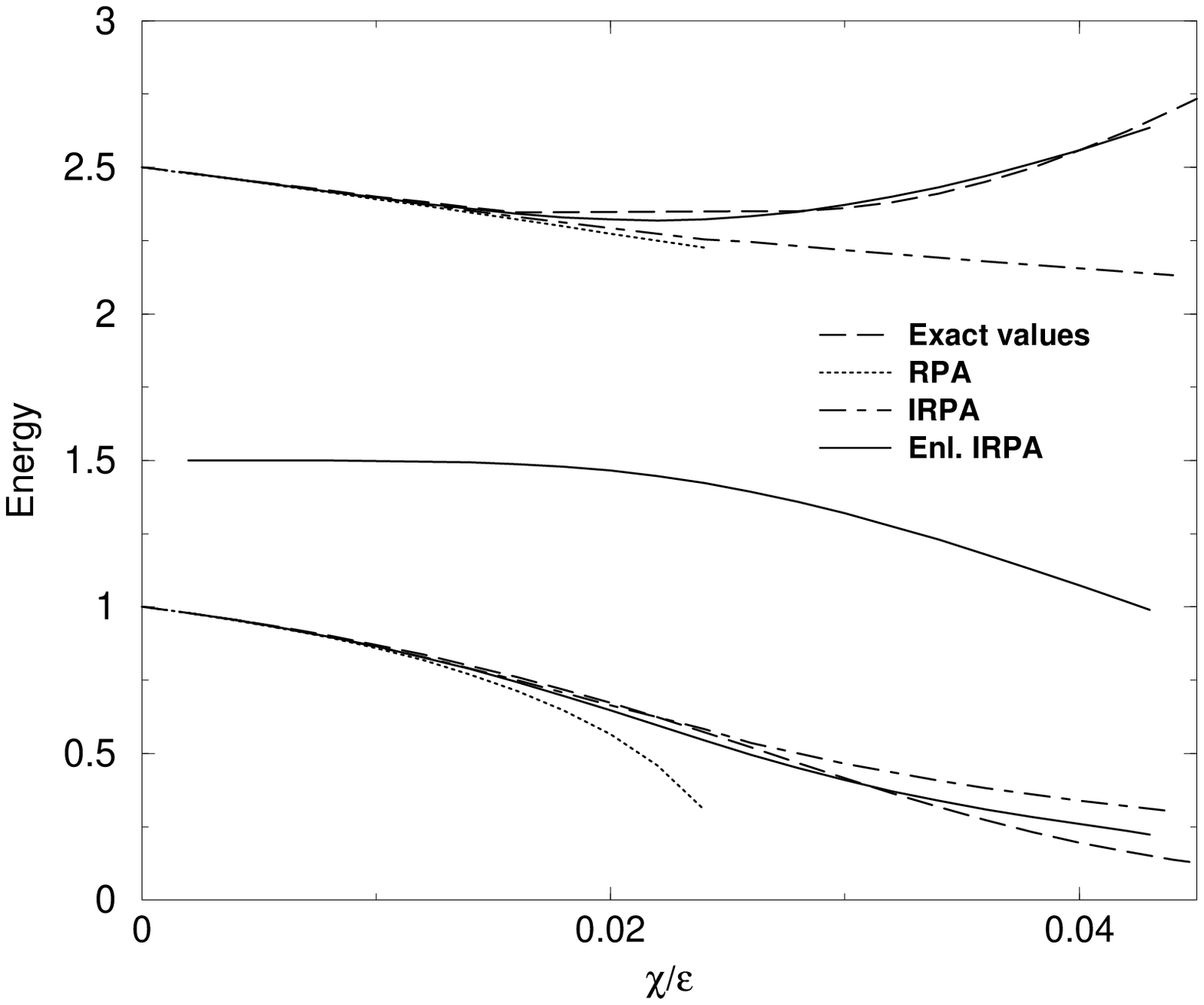}
\mbox{}\\[8cm]

\newpage

\includegraphics{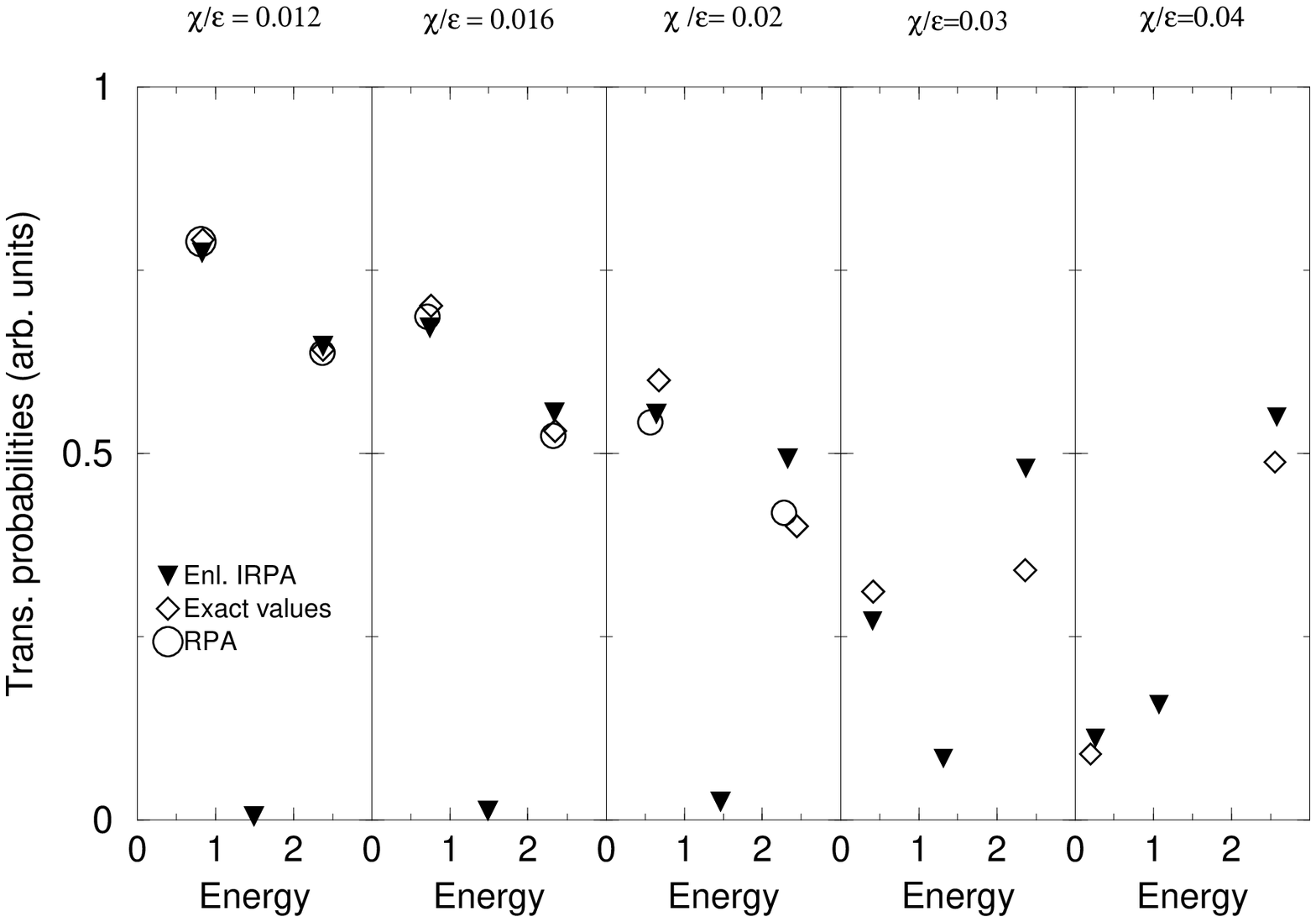}
\mbox{}\\[8cm]

\newpage

\begin{table}
\begin{center}
\begin{tabular}{|c|c|c|c|} \hline
$\chi / \epsilon$ & l.h.s. IRPA  & l.h.s. Enl. IRPA & r.h.s. \\ \hline
 
0.012 & 1.84957  & 2.1877893606 & 2.1877893605 \\ \hline  
0.03 & 0.89411  & 1.3607871867 & 1.3607871868 \\ \hline
0.04 & 0.67858  & 1.5861229231 & 1.5861229230 \\ \hline
\end{tabular}
\caption{The l.h.s of $EWSR$ in the $IRPA$ and in the enlarged $IRPA$ cases, together with the r.h.s., for different values of $\chi / \epsilon$
 and with the parameters (\ref{33})-(\ref{34}).}
\end{center}
\end{table}
  
\end{document}